\documentclass{aa}  

\usepackage{graphicx}
\usepackage[dvipsnames]{xcolor}
\usepackage{multirow}
\usepackage{amsmath}

\newcommand{\kms}{km~s$^{-1}$}
\newcommand{\fluxcgs}{erg~s$^{-1}$~cm$^{-2}$}

\usepackage{txfonts}
\usepackage[]{hyperref}
\hypersetup{
    colorlinks=true,
    citecolor=blue,
    linkcolor=blue,
    urlcolor=blue,
    }
\newcommand{\orcid}[1]{\href{https://orcid.org/#1}{\includegraphics[width=10pt]{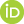}}}

\begin{document}

   \title{The nebular phase of SN 2024ggi: a low-mass progenitor with no signs of interaction}

   \author{L. Ferrari
          \inst{1,2}
          \thanks{luciaferrari@fcaglp.unlp.edu.ar}
          \orcid{0009-0000-6303-4169}
          \and
          G. Folatelli\inst{1,2,3}\orcid{0000-0001-5247-1486}
          \and
          K. Ertini\inst{1,2}\orcid{0000-0001-7251-8368}
          \and
          H. Kuncarayakti \inst{4,5}\orcid{0000-0002-1132-1366}
          \and
          T. Regna \inst{1,2}\orcid{0009-0006-9814-4468}
          \and
          M. C. Bersten \inst{1,2,3}\orcid{0000-0002-6991-0550}
          \and
          C. Ashall \inst{6}\orcid{0000-0002-5221-7557}
          \and
          E. Baron \inst{7,8,9}\orcid{0000-0001-5393-1608}
          \and
          C. R. Burns \inst{10}\orcid{0000-0003-4625-6629}
          \and
          L. Galbany \inst{11,12}\orcid{0000-0002-1296-6887}
          \and
          W. B. Hoogendam \inst{6}\orcid{0000-0003-3953-9532}
          \and
          K. Maeda \inst{13}\orcid{0000-0003-2611-7269}
          \and
          K. Medler \inst{6}\orcid{0000-0001-7186-105X}
          \and
          N. I. Morrell \inst{14}\orcid{0000-0003-2535-3091}
          \and
          B. Shappee \inst{6}\orcid{0000-0003-4631-1149}
          \and
          M. D. Stritzinger \inst{15}\orcid{0000-0002-5571-1833}
          \and
          H. Xiao \inst{16}\orcid{0009-0006-9436-7197}
          }

   \institute{Instituto de Astrofísica de La Plata (UNLP - CONICET), Paseo del Bosque S/N, 1900, Buenos Aires, Argentina
         \and
             Facultad de Ciencias Astronómicas y Geofísicas, Universidad Nacional de La Plata, Paseo del Bosque S/N B1900FWA, La Plata, Argentina
        \and
            Kavli Institute for the Physics and Mathematics of the Universe (WPI), The University of Tokyo, Kashiwa, 277-8583 Chiba, Japan
        \and
            Finnish Centre for Astronomy with ESO (FINCA), 20014 University of Turku, Finland
        \and
            Tuorla Observatory, Department of Physics and Astronomy, 20014 University of Turku, Finland
        \and
            Institute for Astronomy, University of Hawai'i at Manoa, 2680 Woodlawn Dr., Hawai'i, HI 96822, USA
        \and
            Planetary Science Institute, 1700 E. Fort Lowell Rd, Ste 106, Tucson, AZ 85719 USA \email{ebaron@psi.edu}
        \and
            Hamburger Sternwarte, Gojenbergsweg 112, 21029 Hamburg, Germany
        \and
            Homer L.~Dodge Dept.~of Physics and Astronomy, University of Oklahoma, 440 W.  Brooks, Rm 100, Norman, OK 73019 USA
        \and
            Observatories of the Carnegie Institution for Science, 813 Santa Barbara Street, Pasadena CA, 91104, USA
        \and
            Institute of Space Sciences (ICE-CSIC), Campus UAB, Carrer de Can Magrans, s/n, E-08193 Barcelona, Spain
        \and
            Institut d'Estudis Espacials de Catalunya (IEEC), 08860 Castelldefels (Barcelona), Spain
        \and
            Department of Astronomy, Kyoto University, Kitashirakawa-Oiwake-cho, Sakyo-ku, Kyoto 606-8502, Japan
        \and
            Las Campanas Observatory, Carnegie Observatories, Casilla 601, La Serena, Chile
        \and
            Department of Physics and Astronomy, Aarhus University, Ny Munkegade 120, DK-8000 Aarhus C, Denmark %max
        \and
            Department of Physics, Florida State University, 77 Chieftan Way, Tallahassee, FL 32306, USA
            }

   \date{Received Month XX, XXXX; accepted Month XX, XXXX}

% \abstract{}{}{}{}{} 
% 5 {} token are mandatory
 
  \abstract
   {SN 2024ggi is a Type II supernova (SN) discovered in the nearby galaxy NGC 3621 (D $\approx6.7\pm0.4$ Mpc) on 2024 April 03.21 UT. Its proximity enabled detailed investigation of the SN’s properties and its progenitor star. This work focuses on the optical evolution of SN 2024ggi at the nebular phase.}
   {We investigate the progenitor properties and possible asymmetries in the ejecta by studying the nebular phase evolution between days 287 and 400 after the explosion.}
   {We present optical photometry and spectroscopy of SN 2024ggi during the nebular phase, obtained with the Las Campanas and Gemini South Observatories. Four nebular spectra were taken at 287, 288, 360, and 396 days post-explosion, supplemented by late-time $uBVgri$-band photometry spanning $320–400$ days. The analysis of the nebular emission features is performed to probe ejecta asymmetries. Based on the [O I] flux and [O I]/[Ca II] ratio, and comparisons with spectra models from the literature, we arrive to an estimate of the progenitor mass. Additionally, we construct the bolometric light curve from optical photometry and near-infrared data to derive the synthesized nickel mass.}
   {Our analysis suggests a progenitor zero-age-main-sequence mass between $12-15~M_\odot$. The late-time bolometric light curve is consistent with a synthesized $^{56}$Ni mass of $0.05-0.06~M_\odot$. The line profiles exhibit only minor changes over the observed period, and suggest a roughly symmetrical ejecta, with a possible clump of oxygen-rich material moving towards the observer. No signatures of circumstellar material interaction are detected up to 400 days after the explosion.}
   {}

   \keywords{supernovae:general --- 
supernovae:individual:SN 2024ggi}

   \maketitle
%
%-------------------------------------------------------------------

\section{Introduction}\label{sec:intro}

% Core-collapse supernovae (SNe) are the result of the final evolutionary stages of massive stars, i.e., stars with initial masses $\gtrsim 8~\mathrm{M}_\odot$. Most of these stellar explosions correspond to Type II SNe \citep{Li2011}, which have retained their hydrogen-rich envelope and are believed to arise from red supergiant (RSG) stars, as proven by the progenitor identification on archival images \citep{VanDyk2003,maund2005}. The mass of the progenitor star in the zero-age main sequence (ZAMS) plays a major role in the stars' evolution and subsequent explosion features. Their study has deepened during the last decades, where the increasing amount of data has allowed astronomers to better determine the explosion characteristics and the progenitor's properties. 
% Nevertheless, there is a significant discrepancy in the higher end of the mass distributions that come from RSG observations and SNe progenitors, referred by \citealt{Smartt2009} as the \textit{RSG problem}. In this context, it becomes crucial to ...

% ---
Core-collapse supernovae (SNe) mark the demise of massive stars with initial masses $\gtrsim 8~\mathrm{M}_\odot$. The majority of these explosions are classified as hydrogen-rich Type II SNe \citep{Li2011}. They are believed to arise from stars in the red supergiant (RSG) phase that managed to retain a substantial hydrogen-rich envelope. 
Observationally, the association with RSG stars has been supported by direct progenitor identifications in archival images \citep{VanDyk2003, maund2005}. 
In spite of the well-established connection with RSGs, there are major uncertainties about their initial masses and evolutionary paths.
The zero-age main sequence (ZAMS) mass of the progenitor star plays a decisive role in governing its evolutionary path and the resulting SN observational features. 
% Over the past several decades, advances in transient surveys and progenitor studies have significantly refined our understanding of these explosions.

% A persistent and unresolved discrepancy exists at both the low-mass and the high-mass end of the progenitor mass distribution.

Despite the efforts from observational and theoretical works, there are large unknowns regarding the progenitor mass distribution of SNe.
On one hand, it is uncertain which is the lower limiting mass for core-collapse, and this limit value varies depending on the numerical modeling and physical parameters of the star, such as metallicity, rotation, mass-loss rates, and mixing processes \citep{Eldridge2004,Ibeling2013}.
On the other hand, 
while stellar evolution models predict that stars up to $\sim 25~M_\odot$ should undergo successful core-collapse explosions, pre-SN observations of RSGs indicate an apparent upper mass limit of only $\sim 17~{M}_\odot$ \citep{Smartt2009}.
This tension, known as the \textit{RSG problem}, suggests either (1) a fundamental gap in our understanding of late-stage massive stellar evolution (e.g., enhanced mass loss \citealt{Crowther2007}, binary interactions \citealt{Podsiadlowski2004,Smith2014}, or fallback-induced black hole formation \citealt{MacFadyen2001}), or (2) systematic biases in progenitor detection and mass determination.
% In addition to pre-explosion data, LC modeling and comparison between spectral models.  
% Moreover, a recent work from Martinez et al. modelling a large dataset of SNe II show that also an upper limit of XXX was need to explian most of the data, in line with...
In this context, it becomes essential to obtain increasingly precise progenitor mass determinations through both improved pre-explosion imaging and detailed modeling of supernova light curves and spectra, particularly for events occurring in nearby galaxies with well-characterized stellar populations.

Another principal factor governing the evolution of high-mass stars is their mass loss. This mass loss can be driven by line-driven winds, eruptive episodes, and/or binary interactions. These processes not only shape the stellar evolution but also create extensive circumstellar material (CSM) with which the SN ejecta subsequently interacts with which the SN ejecta subsequently interacts. Such interactions manifest observationally through early light curve evolution \citep{Forster2018}, high luminosities and blue colors at early times \citep{Jacobson-Galan2024}, X-ray and radio emission \citep{Chevalier2017,Moriya2021}, highly ionized emission features in early-time spectra (known as flash features; \citealt{Gal-Yam2014,Khazov2016,Yaron2017,Bruch2021,Jacobson-Galan2022}), and through strong, boxy emission profiles in late-time spectra \citep{Kotak2009,Weil2020,Folatelli2025,Jacobson-galan2025arXiv}.

Constraining both the progenitor masses and their mass-loss histories is therefore essential for understanding the final evolutionary stages and explosion mechanisms of massive stars. The nebular phase, beginning $\sim100 - 150$ days post-explosion when the ejecta become optically thin, provides particularly valuable diagnostics. During this phase, photons from the innermost regions escape, revealing the explosion geometry and nucleosynthetic products. Observed asymmetries in the nebular emission lines may reflect intrinsic asymmetries in the explosion itself, offering clues to the progenitor's structure and explosion dynamics. 
Late-time modeling of SNe spectra has provided another tool to investigate the progenitor properties, and more specifically, the progenitor masses \citep{Jerkstrand2012,Dessart2021}. % for the determination of the progenitor...
Recent work by \citet{Fang2025arxiv} has further addressed the RSG problem by demonstrating that nebular spectral analysis supports the absence of high-mass RSG progenitors, reinforcing the possibility to revise stellar evolution models.

SN 2024ggi, discovered on April 11, 2024, by the Asteroid Terrestrial-impact Last Alert System \citep[ATLAS;][]{Tonry2018} in NGC 3621, represents a pivotal case study. With a first detection at JD = 2460411.64 \citep{Tonry2024,Chen2025} and a rapid classification as a Type II SN \citep{Hoogendam2024,Zhai2024}, it stands alongside SN 2023ixf as one of the nearest and most intensively observed core-collapse SNe in recent years (see e.g., \citealt{jacobsongalan2023,kilpatrick2023,bersten2024,Ferrari2024}).

Its proximity and early discovery enabled intensive monitoring campaigns, resulting in high-cadence ultraviolet, optical, and infrared (IR) observations. The SN presented flash ionized features, as reported by \citet{Hoogendam2024,Zhang2024,Pessi2024,JacobsonGalan2024_ggi} and \citet{Shrestha2024}.
Multiwavelength follow-up efforts yielded X-ray detections within the first three days post-explosion \citep{Margutti2024,Lutovinov2024,Zhang2024} and a radio counterpart weeks later \citep{Ryder2024}, while no significant emission was found in the centimeter/millimeter range \citep{Chandra2024,Hu2025} nor in $\gamma$-rays \citep{Marti-Devesa2024}.
\citet{Baron2025arXiv} reported IR observations obtained with the James Webb Space Telescope (JWST) and ground-based telescopes during the plateau phase, $\sim55$ days after the explosion.

The archival image analyses showed a progenitor candidate consistent with an RSG star.
\citet{Xiang2024} combined Hubble Space Telescope (\textit{HST}) and Spitzer Space Telescope pre-explosion images to characterize the progenitor as a bright, red star with an estimated initial mass of $13 \pm 1~M_\odot$. 
\citet{Chen2024}, using Dark Energy Camera Legacy Survey data, independently proposed a slightly more massive progenitor ($14-17$ $M_\odot$).
An alternative approach by \citet{Hong2024} leveraged \textit{HST} imaging to analyze the local stellar environment. By dating the youngest stellar population near the supernova site, they inferred a progenitor mass of 10.2 $M_\odot$, providing a complementary estimate to direct detection methods.
\citet{Ertini2025} presented the first hydrodynamical model of the bolometric light curve during the full plateau phase and derived a progenitor mass of $15~M_\odot$.
Lastly, a recent work by \citet{Dessart2025arXiv_ggi} proposes that the nebular features in the optical and NIR spectra are well reproduced by a progenitor of M = $15.2~M_\odot$.

This paper is organized as follows. Section \ref{sec:observations} presents our observational dataset, including details of the data reduction procedures and the adopted reddening and distance parameters. In Section \ref{sec:spectral_properties}, we carry out a detailed analysis of the nebular-phase spectra, with particular emphasis on emission line profiles and their implications for ejecta geometry. Section \ref{sec:LC} examines the late-time light curve evolution and provides estimates of the synthesized $^{56}$Ni mass. We then derive progenitor mass constraints from nebular spectra in Section \ref{sec:prog_mass}. Finally, Section \ref{sec:conclusion} synthesizes our key findings and discusses their implications, placing SN~2024ggi in context with both previous progenitor studies and our current understanding of Type II SNe.

\section{Observations, reddening and distance}\label{sec:observations}
This work is based on spectroscopic and photometric data acquired at the Gemini South Observatory and the Las Campanas Observatory. 
The spectra acquired at the Las Campanas Observatory were taken by the Precision Observations of Infant Supernova Explosions (POISE\footnote{\url{https://poise.obs.carnegiescience.edu/}}, \citealt{Burns2021}) collaboration.
Phases presented in this paper are referred to an explosion date of JD = $2460411.295 \pm 
0.345$, which corresponds to the midpoint between the first detection of the SN \citep[JD = 2460411.64, ][]{Tonry2024} and the last non-detection \citep[JD = 2460410.95, ][]{Killestein2024}.
The heliocentric redshift of the host galaxy reported by \citet{Koribalski2004} is $\mathrm{z_{Hel}}=0.0024$ and it is used throughout the work to correct the spectra and the light curve phases. 

The distance to NGC 3621 was measured by \citet{Paturel2002} using the Cepheid period-luminosity relation, yielding D~$= 6.7\pm0.5$ Mpc. An independent determination by \citet{Saha2006} obtained D~$=7.4\pm0.4$ Mpc through a similar methodology. Given the comparable uncertainties between the two measurements, we adopt the former value to maintain consistency with the analysis of \citet{Ertini2025}, facilitating direct comparison of results.

Several independent studies have estimated the host galaxy extinction for this source using different methods. 
Following the calibration from \citet{Stritzinger2018}, \citet{JacobsonGalan2024_ggi} derived $E(B-V)_{host} = 0.084 \pm 0.018$ mag by analyzing the Na I D1 and D2 equivalent widths in high-resolution spectra. 
In contrast, \citet{Pessi2024} and \citet{Shrestha2024} obtained lower values of $E(B-V)_{host}=0.036\pm0.007$ mag and $0.034\pm0.020$ mag, respectively, using the relations from \citet{Poznanski2012}. 
When combined with the Milky Way's contribution, the total extinction estimates from these studies are consistent due to a different treatment of the Milky Way extinction contribution. \citet{JacobsonGalan2024_ggi} report $E(B-V)_{tot}=0.154$ mag, \citet{Pessi2024} find 0.16 mag, and \citet{Shrestha2024} measure 0.154 mag. 
Given the uncertainties in these values, we adopt the coincident total reddening of  $E(B-V)_{tot} = 0.16$ mag for our analysis using the \citet{Cardelli1989} law with ${R_v}=3.1$.

\subsection{Spectra}

Two optical spectra were taken with the Inamori-Magellan Areal Camera \& Spectrograph (IMACS, \citealt{Dressler2011}), mounted at the 6.5 m Baade telescope, on 2025 January 23 and 2025 April 06 UT, 286.8 and 359.6 rest-frame days relative to the explosion date given above. All spectra were acquired with the grism Gri-300-17.5; the first one consisted of two 600-s exposures, while the last one consisted of two 900-s exposures. The dispersion of both spectra is $\approx5$ \AA, thus the spectral resolution is $\approx 250$ \kms\ at 6000 \AA.

Another spectrum was obtained with the Gemini Multi-Object Spectrograph \citep[GMOS;][]{hook2004} mounted on the Gemini South Telescope (Program GS-2025A-Q-120, PI Ferrari) on 2025 January 24 UT, 287.7 rest-frame days after the explosion.
The observations were divided into four $900$-s exposures in long-slit mode with the R400 grating. The data were processed with standard procedures using the IRAF Gemini package. Flux calibration was performed with a baseline standard observation.
The spectrum dispersion is $\approx 10$ \AA, resulting in a spectral resolution of $\approx 500$ km s$^{-1}$ at $6000$ \AA. 
% R = 6000 A / 10 A = c / dv → dv = 500 km/s

The last epoch was taken on 2025 May 12 and 13 UT, with the Low Dispersion Survey Spectrograph (LDSS3, upgraded from LDSS2, \citealt{Allington-Smith1994}) mounted on the 6.5 m Clay Telescope of the Las Campanas Observatory. We combined both spectra and obtained a single epoch, which, on average, is 395.9 rest-frame days after the explosion date. The grism used was the VPH-ALL, with a slit of 1 arcsec. Both observations consisted of four 300-s exposures each. The dispersion is $\approx 8.5$, which represents a spectral resolution of $\approx 400$ \kms\ at 6000 \AA.

All four spectra are plotted in Fig.~\ref{fig:ggi_profiles}, top panel, normalized by the maximum in the [\ion{O}{i}]$~\lambda\lambda6300,6364$ region.

\subsection{Light curves}
The photometry presented in this work was acquired with the 1-m Henrietta Swope Telescope of the Las Campanas Observatory.
The follow-up spans between February 25th and April 26th, 2025, i.e., roughly between 320 and 400 d in the \textit{BVgri} filters, and until 377 d in the $u$~band. 
The temporal cadence varies with time between 1 to 14 days for the later observations.
% Light curves cover the \textit{uBVgri} filters with a temporal cadence within 1-15 days. 
The reduction was performed with the POISE photometric pipeline, which is based on the Carnegie Supernova Project (CSP) pipeline described in \citet{Contreras2010} and \citet{Krisciunas2017}. The resulting photometry calibrated to the natural system of the Swope Telescope is listed in Table \ref{tab:photometry} and shown in Fig.~\ref{fig:LC}.
% \textbf{[Add Swope photometry details.]}

The Gemini Program GS-2025A-Q-120 included three 30-s exposures in each $g'$ and $r'$ bands, which were used to calibrate the spectrum. These provided two additional points to the light curve dated January 24th, 2025, 288.4 days after the explosion. The images were reduced using the Dragons software\footnote{\url{https://github.com/GeminiDRSoftware/DRAGONS}} \citep{Labrie2023}. 
We performed the photometry measurements of the SN using the Autophot Python package \citep{Brennan2022}, which employs point-spread function (PSF) photometry for sources with signal-to-noise ratio (SNR) above our detection threshold of 25. The instrumental magnitudes were calibrated to the standard Sloan system using zero points derived from field stars from the Pan-STARRS catalog \citep{Flewelling2020}.

% The photometry of the SN was measured using the Autophot Python package \citep{Brennan2022}, which performs point-spread function (PSF) photometry if the signal-to-noise ratio (SNR) of the source is above an umbral, in this case set to 25. The instrumental magnitudes are then calibrated to the standard system estimating a zero point, which is calculated using $\sim$20 Pan-STARRS field stars \citep{Flewelling2020}. 

We have included data from the Asteroid Terrestrial-impact Last Alert System \citep[ATLAS;][]{Tonry2018} downloaded from its public forced photometry service (\url{https://fallingstar-data.com/}, \citealt{Smith2020}). 

\section{Spectral properties} \label{sec:spectral_properties}

\begin{figure*}
    \centering
    \includegraphics[width=\linewidth]{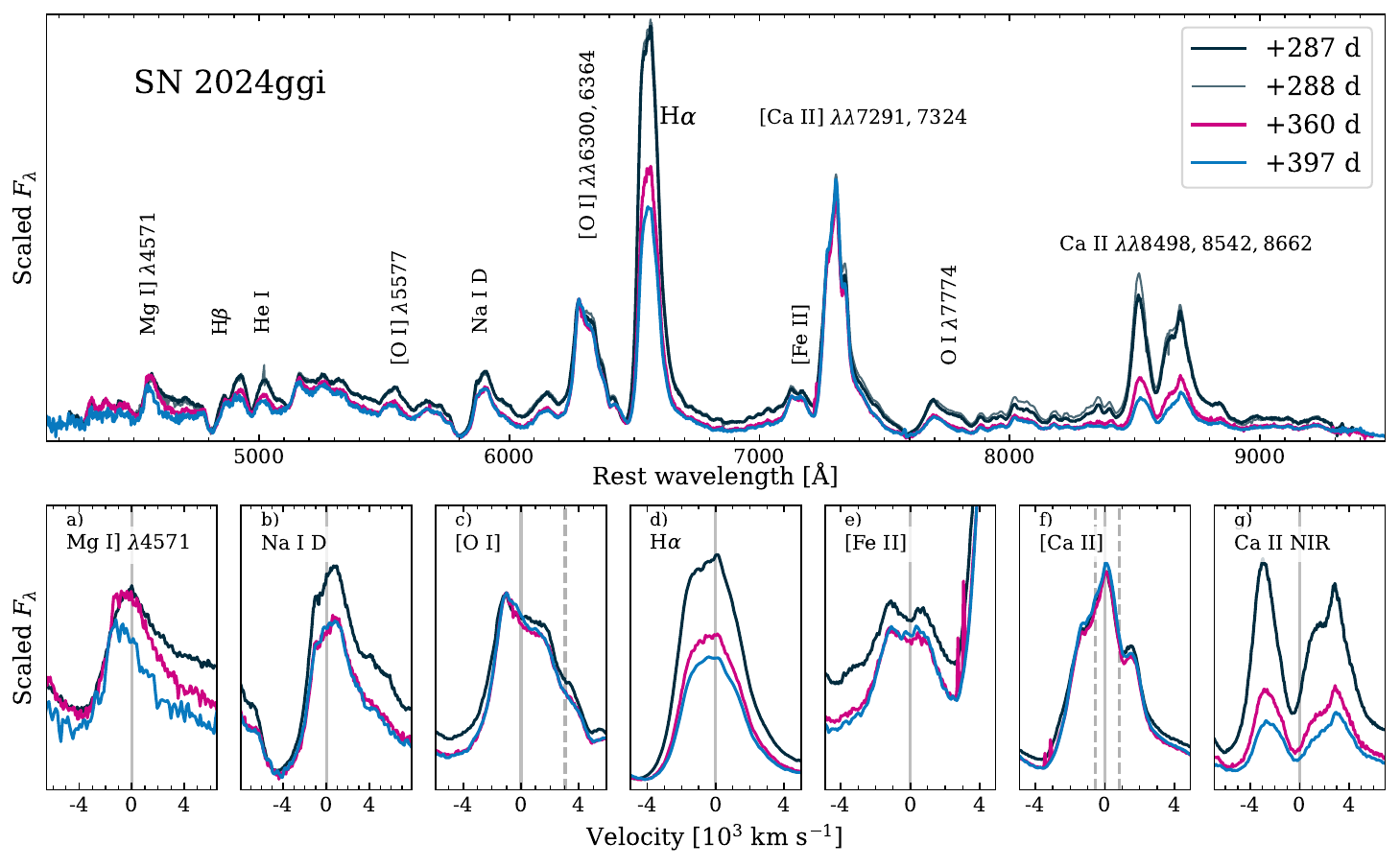}
    \caption{Top panel: SN 2024ggi nebular spectra at 287, 288, 360, and 396 days after the explosion. The 288 d (GMOS) spectrum is plotted with a thin line as it is almost the same epoch as the 287 d (IMACS) spectrum (we do not combine them as they were taken with different instruments). Bottom panels: Line profiles at 287, 360, and 396 days in the velocity space for a) \ion{Mg}{i}]$\lambda4571$, b) \ion{Na}{i} D, c) [\ion{O}{i}]$\lambda\lambda6300,6364$, d) H$\alpha$, e) [\ion{Fe}{ii}]$\lambda7155$, f) [\ion{Ca}{ii}]$\lambda\lambda7291,7324$, g) \ion{Ca}{ii} T. For better display, all spectra are scaled to the maximum emission in the [\ion{O}{i}] region.}% In panel c) the rest }
    \label{fig:ggi_profiles}
\end{figure*}

\begin{figure*}[!ht]
    \centering
    \includegraphics[width=\linewidth]{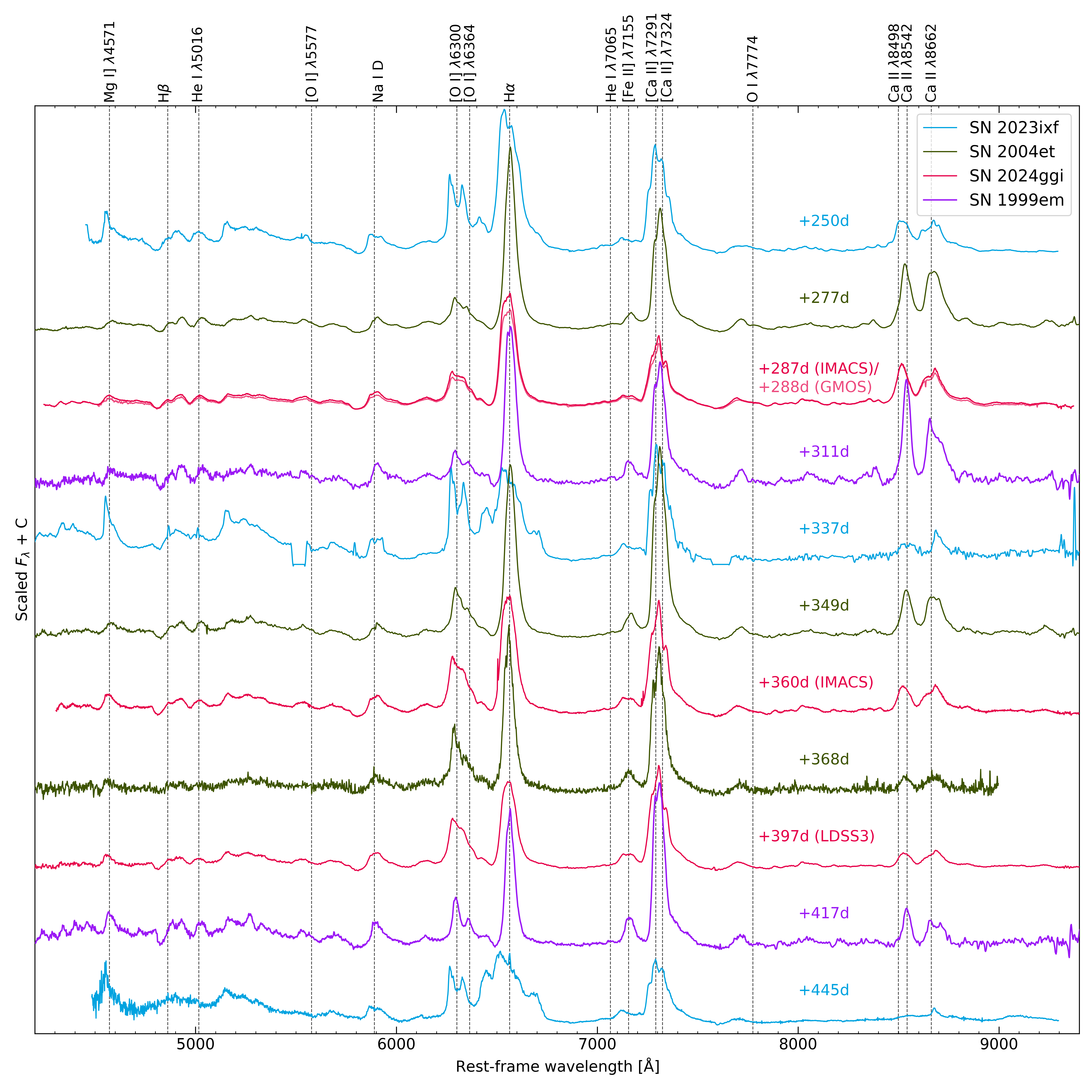}
    \caption{Nebular spectra of SN~2024ggi compared with SN~2023ixf, SN~2004et and SN~199em. All spectra have been corrected by their reported redshift and scaled by the mean value in the region between $7600-8400$ \AA. Emission lines are marked in gray dotted lines.}
    \label{fig:spec_comp}
\end{figure*}

Fig.~\ref{fig:ggi_profiles} presents the spectrum obtained with GMOS, the two IMACS spectra, and the LDSS3 spectrum. The lower panels illustrate the dominant features in velocity space. Over the time interval covered by the observations, the supernova exhibits slow evolution, with the line profiles retaining their shapes while gradually declining in flux. In Fig.~\ref{fig:spec_comp}, we compare these spectra with those of other well-observed Type II SNe during the nebular phase. The spectra at 287 and 288 days are overplotted, as they are nearly indistinguishable. The spectra of SN 2024ggi display the characteristic emission lines expected for a Type II SN in the nebular phase, with the primary emissions arising from oxygen, hydrogen, calcium, iron, and sodium ions.

The strongest emission lines are centered at their rest wavelengths. Notably, H$\alpha$ remains nearly at rest, and its flux diminishes more rapidly than that of other lines. A detailed analysis of the [\ion{O}{i}] and H$\alpha$ features is provided in Sect.~\ref{sec:line_profiles}.

When compared to the extensive sample of Type II SNe from \cite{Silverman2017}, the nebular spectra of SN 2024ggi appear typical, though with some subtle distinctions. The H$\alpha$ profile is marginally broader than usual (see Sect.~\ref{sec:line_profiles}),  and the [O I] doublet exhibits a persistent single-peaked profile throughout the observed period.
% in contrast to the double-peaked structure often observed. 
This profile likely reflects the spatial distribution of the ejecta, as discussed further in Sect.~\ref{sec:line_profiles}.

Finally, we note the lack of an emission feature between [O I] and H$\alpha$—detected in SN 2023ixf (Fig.~\ref{fig:spec_comp}) and attributed to relatively early CSM interaction \citep{Kumar2025,Folatelli2025}.
This absence suggests that, contrary to SN 2023ixf, SN 2024ggi did not experience interaction between rapidly expanding SN ejecta with circumstellar material.
However, extended spectroscopic monitoring beyond 600–700 days will be essential to determine whether late-phase interaction features emerge, as observed in some normal Type II SNe \citep{Kotak2009,Weil2020}.

% The most prominent emission lines appear centered at the rest wavelength.
% H$\alpha$ is centered nearly at rest, and its flux decreases faster compared to other lines.
% See Sect.~\ref{sec:line_profiles} for a detailed analysis of the [\ion{O}{i}] and the H$\alpha$ features.

% The nebular spectra of SN 2024ggi appear normal when compared with a large sample of SNe II from \cite{Silverman2017}. However, some slight differences can be noticed. The H$\alpha$ profile is slightly wider than usual (see Sect.~\ref{sec:line_profiles}). Another noticeable difference resides in the [O I] doublet profile, which is single-peaked for SN 2024ggi. This is related to the spatial distribution of the material; see further discussion in Sect.~\ref{sec:line_profiles}. 

%Explosion time of 99em JD= 2451475.6 from \citet{leonard2002}.

% \begin{figure}
%     \centering
%     \includegraphics[width=0.99\linewidth]{figures/24ggi_23ixf.png}
%     \caption{Comparison of SN 2024ggi with other typical type II SNe.}
%     \label{fig:comparison}
% \end{figure}
\subsection{Line profiles and ejecta asymmetries}
\label{sec:line_profiles}
% In this Section, we explore the detailed profiles of the [\ion{O}{i}]$~\lambda\lambda6300,6364$, H$\alpha$, and [\ion{Ca}{ii}]$~\lambda\lambda7291,7324$ emissions with the aim of unveiling the structure of the SN ejecta. For this purpose, we used the IMACS spectra at 287 and 360 days, and the LDSS3 spectrum at 396 days, all corrected by the redshift and extinction values detailed in Sect.~\ref{sec:observations}.

% In this section, we analyze the profiles of the main emission lines and perform Gaussian decomposition of the [\ion{O}{i}] and H$\alpha$ features.
The \ion{Mg}{i}] $\lambda$4571 emission exhibits an asymmetric profile, with a peak that gradually evolves toward the blue, and a weakening of the red wing (see Fig.~\ref{fig:ggi_profiles}, panel a).
The Na I D line maintains a residual P~Cygni profile with noticeable asymmetry (Fig.~\ref{fig:ggi_profiles}, panel b). This line might suffer from contamination due to \ion{He}{i} $\lambda5876$.

\begin{figure}
    \centering
    \includegraphics[width=0.9\linewidth]{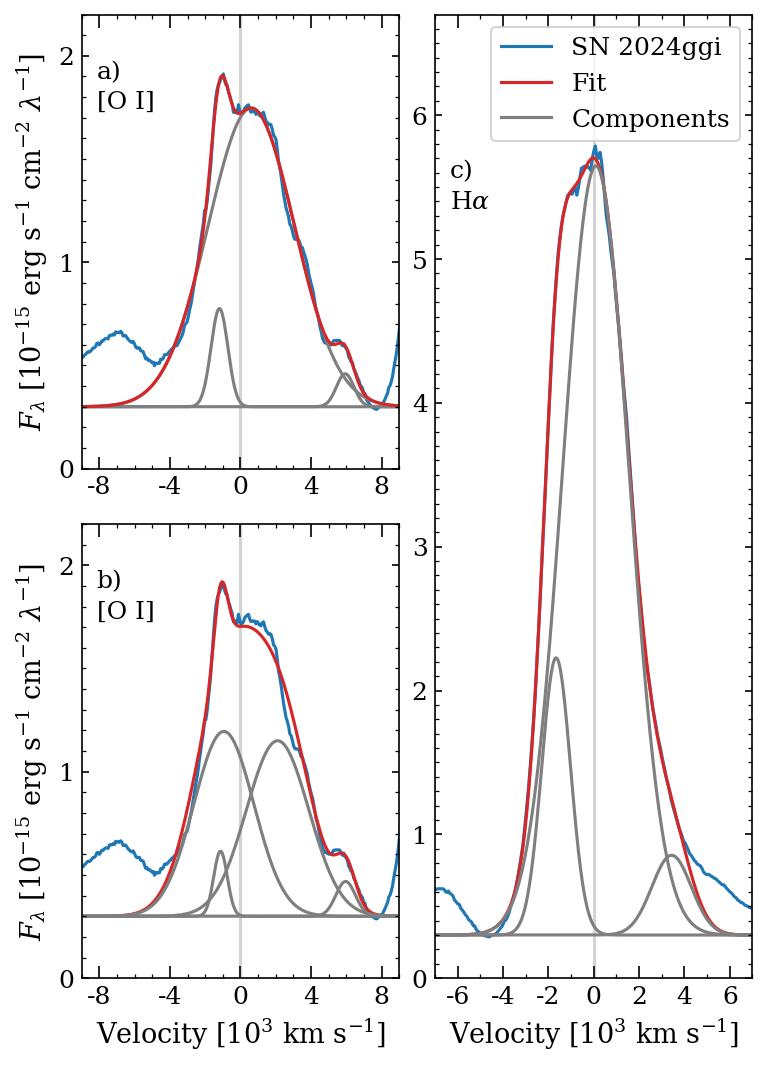}
    \caption{Gaussian decomposition of the [\ion{O}{i}]$\lambda\lambda6300,6364$ and H$\alpha$ line profiles in the spectrum of SN 2024ggi at 287 days. Panel a): three-Gaussian fitting of the [\ion{O}{i}] doublet, with a wide component centered nearly at the rest wavelength (i.e., 6300 \AA). Panel b): four-Gaussian fitting of the [\ion{O}{i}] doublet with the separation of the two wide emissions fixed at 64 \AA. Panel c): three-Gaussian fitting of the H$\alpha$ line, resulting in a wide component centered at rest. In Table \ref{tab:fitting}, all FWHM and peak velocities are detailed for the three fittings.}
    \label{fig:fitting}
\end{figure}

The [\ion{O}{i}] $\lambda\lambda$6300,6364 doublet displays a persistent single-peaked profile throughout our observations, accompanied by a narrow emission component blue-shifted by $\approx-1200$ km s$^{-1}$ (see Fig.~\ref{fig:ggi_profiles}, panel c). While the blue shift gradually decreases with time, the full width at half maximum (FWHM) of the broad component remains constant.
Fig.~\ref{fig:ggi_profiles} (panel c) demonstrates that the [\ion{O}{i}] profile maintains its complex morphology throughout our observations. We present detailed Gaussian fits to the 287 days spectrum in Fig.~\ref{fig:fitting}, with corresponding parameters listed in Table \ref{tab:fitting}. 
Panel a) depicts a three-component fitting, consisting of a broad central emission (FWHM $\approx$5000 km s$^{-1}$ blue shifted by $\approx$600 km s$^{-1}$) combined with a narrow component (FWHM $\approx$1400 km s$^{-1}$) that evolves from $-1300$ to $-1100$~km s$^{-1}$ (although this difference lies within our spectral resolution; see Sect.~\ref{sec:observations}). An additional component accounts for the red shoulder, which may include contamination from a weak H$\alpha$ P~Cygni profile. 
Panel b), on the other hand, shows a four-component alternative, in which we constrained the doublet separation to 64 \AA, and find both primary components blue shifted by $\approx$900 km s$^{-1}$, while the narrow emission and red shoulder features remain consistent with the three-component approach.
However, in this case the ratio [\ion{O}{i}]$\lambda6300$/[\ion{O}{i}]$\lambda6364$ would be $1:1$, which implies an optically thick regime, not expected for the nebular phase.

A single Gaussian does not fit the H$\alpha$ well due to its pronounced asymmetry. 
A single-component fit yields central velocities near rest with FWHM values of 3900, 3800, and 3700~km s$^{-1}$ at 287, 360, and 396 days, respectively. Our preferred three-component model (Fig.~\ref{fig:fitting}, panel c) includes a central emission (FWHM $\approx$3100 km s$^{-1}$) at rest wavelength, a blue shifted component ($\approx$1600 km s$^{-1}$, FWHM $\approx$1500 km s$^{-1}$), and a red component.
While these velocities do not align with [\ion{N}{ii}] doublet positions, we cannot exclude some [\ion{N}{ii}] and Fe II permitted lines contribution. This multi-component structure likely reflects intrinsic line asymmetry rather than distinct emitting regions, possibly indicating a persistent P~Cygni absorption affecting the blue wing.

The [\ion{Fe}{ii}] $\lambda$7155 feature is particularly noteworthy because it has a stable double-peaked profile at all epochs, with velocity components at $\approx$1000 km s$^{-1}$ and $-1200$ km s$^{-1}$ (see Fig.~\ref{fig:ggi_profiles}, panel e). This distinctive structure is absent in our comparison sample (Fig.~\ref{fig:spec_comp}) and, to our knowledge, has not been previously reported in Type II SNe. The uniqueness of the profile may reflect either an asymmetric distribution of iron-rich ejecta or a self-absorption effect.

The [\ion{Ca}{ii}] profile shows three persistent peaks at 7280 \AA, 7310 \AA, and 7340 \AA~(see Fig.~\ref{fig:ggi_profiles}, panel f). Interpreting the two redder features as doublet components implies a $\approx$500 km s$^{-1}$ blue shift. Alternatively, using the mean doublet wavelength (7307 \AA) as reference would imply a central component near rest and blue and red components shifted by $-1000$ km s$^{-1}$ and 1500 km s$^{-1}$, respectively.
% The [\ion{Ca}{ii}] $\lambda\lambda$7291,7324 doublet is centered at rest wavelength and reveals a complex profile with distinct shoulders at $\approx1500$ km s$^{-1}$ and $\approx$2000 km s$^{-1}$. 
Notably, the \ion{Ca}{ii} NIR triplet fades more rapidly than other lines, with the redder component decaying slightly slower than its bluer counterparts (see Fig.~\ref{fig:ggi_profiles}, panel g).

The overall line profiles suggest that the ejecta is relatively symmetric. The notable exception is the blue-shifted [\ion{O}{i}] component, which may indicate a localized clump of oxygen-rich material moving toward the observer.
The fact that the FWHMs of the [\ion{O}{i}] and the H$\alpha$ lines are similar (see Table \ref{tab:fitting}) may indicate substantial mixing of the H-rich envelope with the core material.

\begin{table}[h!]
\caption{Fitting parameters of the individual Gaussian components depicted in Fig.~\ref{fig:fitting}.}
    \begin{tabular}{c|c|c|c|c}
        Line      & Ref. wavelength & Peak velocity & FWHM & Fig. \ref{fig:fitting}  \\
        &[\AA]&[\kms]&[\kms]&panel\\
        \hline
        {[}O I{]} & 6300           & $-1200$         & 1100 & a)       \\
        {[}O I{]} & 6300           & $600$           & 5700 & a)       \\
        \hline
        {[}O I{]} & 6300           & $-1200$         & 900  & b)       \\
        {[}O I{]} & 6300           & $-900$          & 4000 & b)       \\
        {[}O I{]} & 6364           & $-900$          & 4100 & b)       \\
        \hline
        H$\alpha$ & 6563           & $-1600$         & 1500 & c)       \\
        H$\alpha$ & 6563           & 90            & 3400 & c)      
    \end{tabular}
    \label{tab:fitting}
\end{table}

\section{Late-time light curves and $^{56}$Ni mass}\label{sec:LC}

% The late-time photometric evolution observed with the Henrietta Swope Telescope is presented in Fig.~\ref{fig:LC}, with complete photometry detailed in Table \ref{tab:photometry}. The dataset combines our Swope observations with Gemini \textit{g}- and \textit{r}-band photometry at 288 days, early-phase data from \citet{Ertini2025}, and public ATLAS photometry.

\begin{figure}
    \centering
    \includegraphics[width=\linewidth]{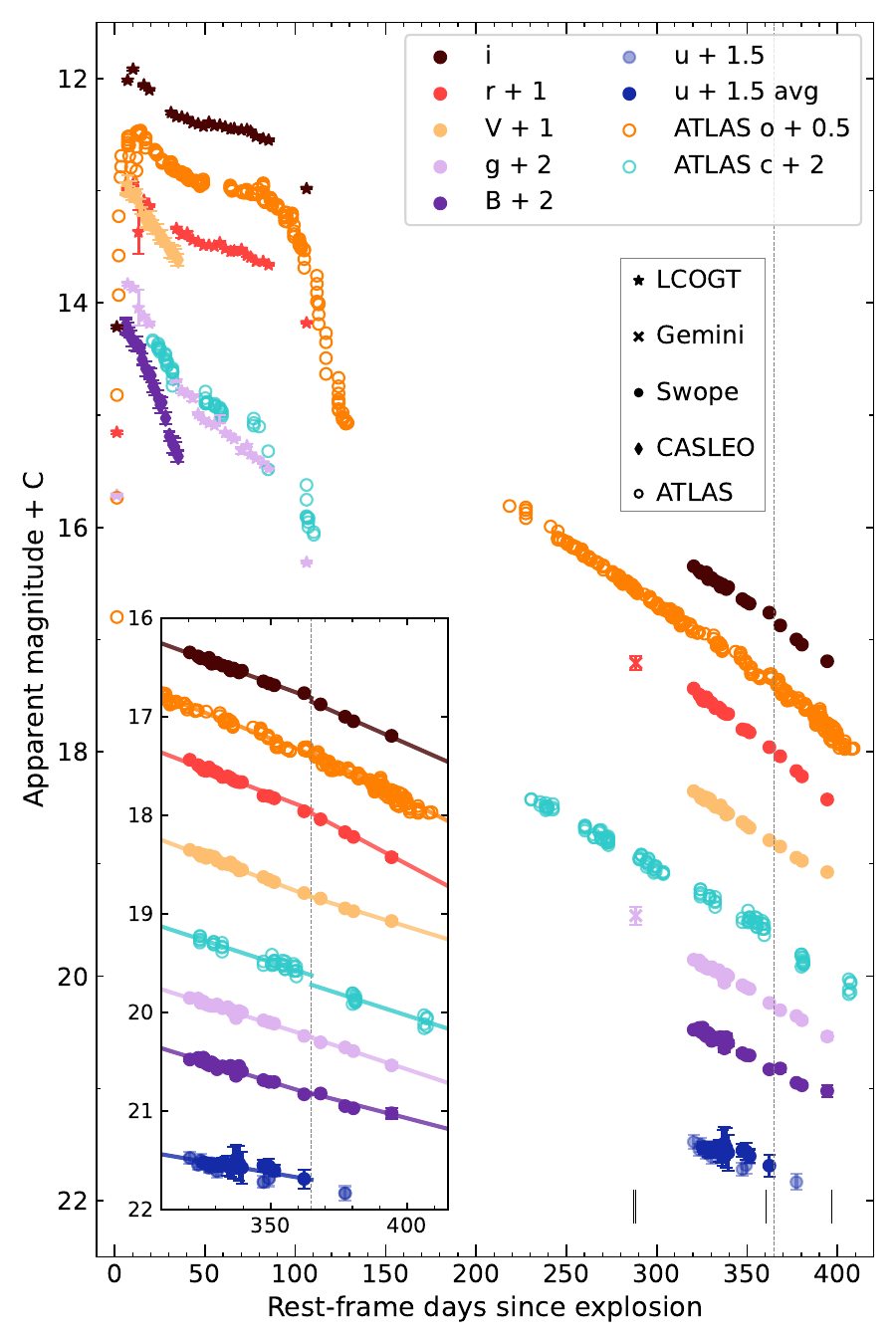}
    \caption{Optical light curves of SN~2024ggi. The new data published in this work correspond to the Gemini and Swope telescopes. Early phase data is published by \citet{Ertini2025}. 
    % ATLAS photometry was downloaded from the public database. 
    The black vertical lines indicate the phases of the four spectra presented in this work. The dotted gray line indicates the epoch where the $o$-, $r$-, and $i$-band light curves show a slope break, also noticeable in the ATLAS o-band light curve (see Sect.~\ref{sec:LC}).
    Inset shows an expanded view of the temporal evolution between $320-415$ days, highlighting the change in decline rate. Solid lines represent separate linear fits to the data before and after 365 days (marked by the vertical dotted line).}
    \label{fig:LC}
\end{figure}

\begin{table*}[h]
    \centering
    \caption{Photometry of the SN 2024ggi spanning from 320 to 380 days after the explosion, obtained with the Henrietta Swope Telescope at Las Campanas Observatory.}
    \begin{tabular}{lcccccccc}
        \hline
        JD &  Phase$^{a}$ &     $u^{b}$  &     $B$ &     $V$ &    $g$ &   $r$ &     $i$  \\
        \hline
        2460731.7 & 319.6 & 19.98 $\pm$  0.06 & 18.48 $\pm$ 0.02 & 17.35 $\pm$ 0.01 & 17.85 $\pm$ 0.01 & 16.44 $\pm$ 0.01 & 16.35 $\pm$ 0.01 \\
        2460734.7 & 322.6 & 20.05 $\pm$  0.05 & 18.46 $\pm$ 0.02 & 17.38 $\pm$ 0.01 & 17.86 $\pm$ 0.01 & 16.49 $\pm$ 0.01 & 16.39 $\pm$ 0.01 \\
        2460735.7 & 323.6 & 20.00 $\pm$  0.06 & 18.50 $\pm$ 0.02 & 17.41 $\pm$ 0.02 & 17.90 $\pm$ 0.02 & 16.51 $\pm$ 0.01 & 16.41 $\pm$ 0.02 \\
        2460736.7 & 324.6 & 20.02 $\pm$  0.04 & 18.46 $\pm$ 0.03 & 17.40 $\pm$ 0.02 & 17.89 $\pm$ 0.02 & 16.54 $\pm$ 0.02 &        $\cdots$       \\
        2460737.7 & 325.6 & 20.03 $\pm$  0.05 & 18.52 $\pm$ 0.02 & 17.43 $\pm$ 0.01 & 17.92 $\pm$ 0.01 & 16.55 $\pm$ 0.01 & 16.42 $\pm$ 0.01 \\
        2460738.7 & 326.6 & 20.08 $\pm$  0.05 & 18.54 $\pm$ 0.02 & 17.41 $\pm$ 0.01 & 17.93 $\pm$ 0.01 & 16.51 $\pm$ 0.01 & 16.40 $\pm$ 0.01 \\
        2460739.7 & 327.6 & 20.03 $\pm$  0.06 & 18.51 $\pm$ 0.02 & 17.42 $\pm$ 0.02 & 17.93 $\pm$ 0.02 & 16.54 $\pm$ 0.01 & 16.46 $\pm$ 0.02 \\
        2460740.7 & 328.6 & 20.04 $\pm$  0.06 & 18.53 $\pm$ 0.03 & 17.42 $\pm$ 0.02 & 17.93 $\pm$ 0.01 & 16.55 $\pm$ 0.01 & 16.45 $\pm$ 0.01 \\
        2460741.7 & 329.6 & 20.11 $\pm$  0.07 & 18.55 $\pm$ 0.04 & 17.46 $\pm$ 0.01 & 17.92 $\pm$ 0.01 & 16.57 $\pm$ 0.01 & 16.45 $\pm$ 0.01 \\
        2460743.7 & 331.6 & 20.05 $\pm$  0.05 & 18.55 $\pm$ 0.02 & 17.49 $\pm$ 0.02 & 17.96 $\pm$ 0.01 & 16.61 $\pm$ 0.01 & 16.48 $\pm$ 0.01 \\
        2460745.6 & 333.5 & 19.97 $\pm$  0.10 & 18.57 $\pm$ 0.02 & 17.50 $\pm$ 0.01 & 17.95 $\pm$ 0.02 & 16.61 $\pm$ 0.01 & 16.49 $\pm$ 0.01 \\
        2460746.6 & 334.5 & 19.93 $\pm$  0.09 & 18.54 $\pm$ 0.02 & 17.48 $\pm$ 0.02 & 17.97 $\pm$ 0.02 & 16.62 $\pm$ 0.02 & 16.52 $\pm$ 0.02 \\
        2460747.7 & 335.6 & 20.24 $\pm$  0.16 & 18.58 $\pm$ 0.03 & 17.52 $\pm$ 0.02 & 18.00 $\pm$ 0.02 & 16.64 $\pm$ 0.02 & 16.51 $\pm$ 0.02 \\
        2460748.6 & 336.5 & 20.17 $\pm$  0.14 & 18.64 $\pm$ 0.04 & 17.52 $\pm$ 0.03 & 18.06 $\pm$ 0.03 & 16.66 $\pm$ 0.02 &        $\cdots$       \\
        2460749.7 & 337.6 & 19.81 $\pm$  0.18 & 18.54 $\pm$ 0.03 & 17.56 $\pm$ 0.02 & 17.98 $\pm$ 0.02 & 16.66 $\pm$ 0.01 & 16.55 $\pm$ 0.01 \\
        2460750.8 & 338.7 & 19.82 $\pm$  0.16 & 18.60 $\pm$ 0.03 & 17.55 $\pm$ 0.01 & 18.00 $\pm$ 0.02 & 16.66 $\pm$ 0.01 & 16.53 $\pm$ 0.01 \\
        2460758.6 & 346.5 & 20.22 $\pm$  0.06 & 18.69 $\pm$ 0.02 & 17.63 $\pm$ 0.01 & 18.08 $\pm$ 0.01 & 16.80 $\pm$ 0.01 & 16.64 $\pm$ 0.01 \\
        2460760.7 & 348.6 & 20.18 $\pm$  0.08 & 18.70 $\pm$ 0.02 & 17.65 $\pm$ 0.01 & 18.10 $\pm$ 0.01 & 16.81 $\pm$ 0.01 & 16.66 $\pm$ 0.01 \\
        2460762.6 & 350.5 & 20.11 $\pm$  0.06 & 18.71 $\pm$ 0.02 & 17.68 $\pm$ 0.01 & 18.11 $\pm$ 0.01 & 16.83 $\pm$ 0.01 & 16.68 $\pm$ 0.01 \\
        2460773.6 & 361.4 & 20.23 $\pm$  0.10 & 18.83 $\pm$ 0.02 & 17.79 $\pm$ 0.02 & 18.24 $\pm$ 0.02 & 16.96 $\pm$ 0.01 & 16.76 $\pm$ 0.02 \\
        2460779.6 & 367.4 & 20.05 $\pm$  0.14 & 18.82 $\pm$ 0.03 & 17.84 $\pm$ 0.02 & 18.30 $\pm$ 0.02 & 17.04 $\pm$ 0.01 & 16.87 $\pm$ 0.01 \\
        2460788.6 & 376.4 & 20.33 $\pm$  0.07 & 18.95 $\pm$ 0.02 & 17.94 $\pm$ 0.01 & 18.35 $\pm$ 0.01 & 17.17 $\pm$ 0.01 & 17.00 $\pm$ 0.01 \\
        2460791.6 & 379.4 &        $\cdots$        & 18.97 $\pm$ 0.02 & 17.97 $\pm$ 0.02 & 18.39 $\pm$ 0.02 & 17.22 $\pm$ 0.01 & 17.05 $\pm$ 0.01 \\
        2460805.6 & 393.3 &        $\cdots$        & 19.02 $\pm$ 0.05 & 18.07 $\pm$ 0.03 & 18.54 $\pm$ 0.03 & 17.42 $\pm$ 0.03 & 17.19 $\pm$ 0.02 \\
        \hline
        \end{tabular}
    \tablefoot{\\
    \tablefoottext{a}{Phases are given in rest-frame days referred to the explosion date, JD = $2460411.3$.}
    \tablefoottext{b}{Magnitudes are presented in the Swope photometric natural system.}
    }
    \label{tab:photometry}
\end{table*}

\begin{table}[]
    \centering
    \caption{Decline rate in all filters for SN 2024ggi in mag/100 d, before and after the break at 365 days.}
    % \begin{tabular}{c|c|c|c|c|c|c}
    %     &u    &B    &V    &     g&r   &i  \\\hline
    %      $t<365$ d&0.42 &0.88 &1.05 &0.88 &1.06 &1.01\\
    %      $t>365$ d& ---  &0.73  &0.86 &0.92  &1.48 &1.22
    % \end{tabular}
    \begin{tabular}{lcc}
    \hline
    Filter &  $t <365$ d &  $t >365$ d  \\
    \hline
     $u$ &           0.47  $\pm$ 0.17 &         --- \\
     $B$ &           0.85  $\pm$ 0.04 &         0.72  $\pm$  0.25 \\
     $V$ &           1.05  $\pm$ 0.03 &         0.86  $\pm$  0.12 \\
     $g$ &           0.88  $\pm$ 0.03 &         0.92  $\pm$  0.15 \\
     $r$ &           1.06  $\pm$ 0.06 &         1.48  $\pm$  0.12 \\
     $i$ &           1.01  $\pm$ 0.03 &         1.22  $\pm$  0.09 \\
    \hline
    $o$ &           1.09  $\pm$ 0.01 &         1.40  $\pm$ 0.06 \\
    $c$ &           0.90  $\pm$ 0.01 &         0.89  $\pm$ 0.11 \\
    \hline
    Bolometric &   1.10 $\pm$ 0.13 & 1.21  $\pm$ 0.31 \\
    \hline
\end{tabular}
    \label{tab:declines}
\end{table}

Between 200 and 400 days, the light curves follow the expected linear decline pattern for Type II SNe. A distinct break appears in the \textit{r}- and \textit{i}-band light curves at approximately 365 days, marking the onset of a slightly faster decline rate. This behavior is reflected in the ATLAS \textit{o}-band data, while the bluer \textit{BVgc} bands (and potentially the \textit{u} band) maintain a nearly constant decline rate throughout the entire observed period (see Fig.~\ref{fig:LC}).

We quantified the decline rates through linear fits to each band, with uncertainties determined via Monte Carlo simulations using 5000 iterations. The resulting measurements, presented in Table \ref{tab:declines}, reveal statistically significant slope changes for the $o$-, $r$-, and $i$-bands following the 365-day break point.
{Before that phase, all values are equal to or lower than that of the decline rate in the case of full $\gamma$-ray trapping, that is, 1 mag per 100 days.}

As we discussed in Sect. \ref{sec:line_profiles}, the H$\alpha$ and \ion{Ca}{ii} NIR triplet fluxes drop faster than other lines such as the [\ion{O}{i}] and [\ion{Ca}{ii}] doublets. 
We tested whether this change in line fluxes could be responsible for the increase in the decline rates seen in the $ori$ bands, as opposed to a change in the continuum shape.
The change in decline rates seen in the $ori$ bands may in part be attributed to a sudden drop in the H$\alpha$ and \ion{Ca}{ii} NIR triplet line fluxes. A rough calculation based on the difference between the 397 and the 287 days spectra scaled to the same epoch by the $^{56}$Co decline rate indicates that the decrease in line fluxes is enough to explain the drop in the $ori$-band fluxes between the extrapolations that result using the pre- and post-356 d decline rates. However, the cause of the sudden change in flux (whether it be line or continuum) remains unclear.
% To assess this hypothesis, we extrapolated the light curves previous and after  365 days and measured the theoretical flux difference at phase 396. 
% We scaled the 287 d spectrum considering a decay rate of 1 mag/100 days, and subtracted the H$\alpha$ profile from the spectrum at 396 days.
% We then measured the synthetic photometry

\subsection{Bolometric luminosity and Ni mass}

During the radioactive tail phase of Type II SNe, the bolometric luminosity is primarily governed by the mass of synthesized $^{56}$Ni. By this stage, hydrogen recombination in the ejecta is complete, and the emission is powered exclusively by the radioactive decay $^{56}$Co $\xrightarrow~ ^{56}$Fe \citep{Anderson2019}. The deposition of energy from $^{56}$Co decay products—$\gamma$ rays and positrons—heats the ejecta and sustains the observed emission. A steeper light-curve decline relative to the instantaneous $^{56}$Co decay rate may indicate partial leakage of gamma rays from the ejecta. % without thermalization.

Additional mechanisms, such as ejecta-CSM interaction, dust formation, or light echoes, could influence the late-time light curve evolution. However, our analysis of the spectral and photometric data up to 400 days post-explosion shows no evidence of substantial ongoing interaction, as demonstrated by the absence of interaction features. 
Similarly, the lack of significant dust formation is inferred from the stability of the emission line profiles. This interpretation is further supported by the absence of dust-related features in late-time IR spectra, as reported by \citet{Dessart2025arXiv_ggi}.
Light echoes, which typically produce a bluer flux excess, would also be expected to flatten the late-time light curve in blue filters. Instead, we observe a continued decline in these bands, ruling out any significant contribution from echoes.
We therefore assume that the late-time emission is dominated by the radioactive decay input, thus neglecting any additional effects.

Our late-time photometry covers the optical wavelength range. At this phase, an important fraction of the total flux from the SN is emitted in the IR range of the electromagnetic spectrum (beyond 8000\AA, see e.g., \citealt{Dessart2025arxiv}).
In order to compute an accurate bolometric flux, we used spectra from the Infrared Telescope Facility (IRTF), dated February 15, 2025 (310 days), and from the JWST, taken on January 21st, 2025 (287 days). These data will be published by Mera et al., in prep; details on the reduction procedure can be found in \citet{Medler25_HISS_arxiv, Hoogendam25_epr_arxiv, Hoogendam25_pxl_arxiv}. 

The IRTF and JWST spectra overlap in the region between 1.7 and 2.3 \textmu m.
Using the absolutely flux-calibrated JWST spectrum as reference, we scaled the IRTF spectrum to match the mean flux level in this overlapping region. This procedure enabled us to construct a continuous, flux-calibrated IR spectrum spanning 0.9-4.1 \textmu m at 287 days.
% Considering the JWST data as absolute-flux calibrated, we scaled the spectrum from IRTF to match the mean value of the overlapping wavelength range. With this, we were able to construct a flux-calibrated IR spectrum covering the range of 0.9 to 4.1 \textmu m at the phase of 287 days.

% Armed with the IRTF+JWST spectrum, we measured the synthetic photometry in the Swope BVgir bands from our IMACS spectrum at 287 d, previously scaled by photometry and corrected by reddening. 
% We then obtained an SED covering the region from 0.44 to 3.5 \textmu m. 
% The total bolometric flux was calculated by integrating the synthetic magnitudes and adding a contribution in the UV and the IR beyond 3.5 \textmu m with a blackbody function fitted to the SED.
% 60\% of the total flux lied outside (50 \% in the IR and 10 \% in the UV) the BVgri bands. 

After obtaining a combined IRTF+JWST spectrum, we derived synthetic photometry in the Swope $BVgri$ bands from our reddening-corrected IMACS spectrum at 287 days (previously flux-scaled using contemporaneous photometry, see Sect.~\ref{sec:observations}). This enabled the construction of a complete spectral energy distribution (SED) spanning 0.44 to 3.5 \textmu m.
We computed the total bolometric flux by integrating the synthetic magnitudes and adding estimated UV ($<0.44$ \textmu m) and far-IR ($>3.5$ \textmu m) contributions derived from a blackbody fit to the SED. Remarkably, 60\% of the total flux originated outside the optical $BVgri$ bands, with 50\% emerging in the IR and 10\% in the UV.

We further assumed that this percentage remained constant throughout the entire tail phase. This assumption is justified by black-body fits to the Swope photometry between 320 and 395 days, which yielded a nearly constant black-body inferred temperature of $T_{BB}=4200-4500$K. Additionally, the slow evolution of optical spectral features suggests that no strong variations in emission lines outside this wavelength range would significantly alter the derived percentages.

\begin{figure}
    \centering
    \includegraphics[width=\linewidth]{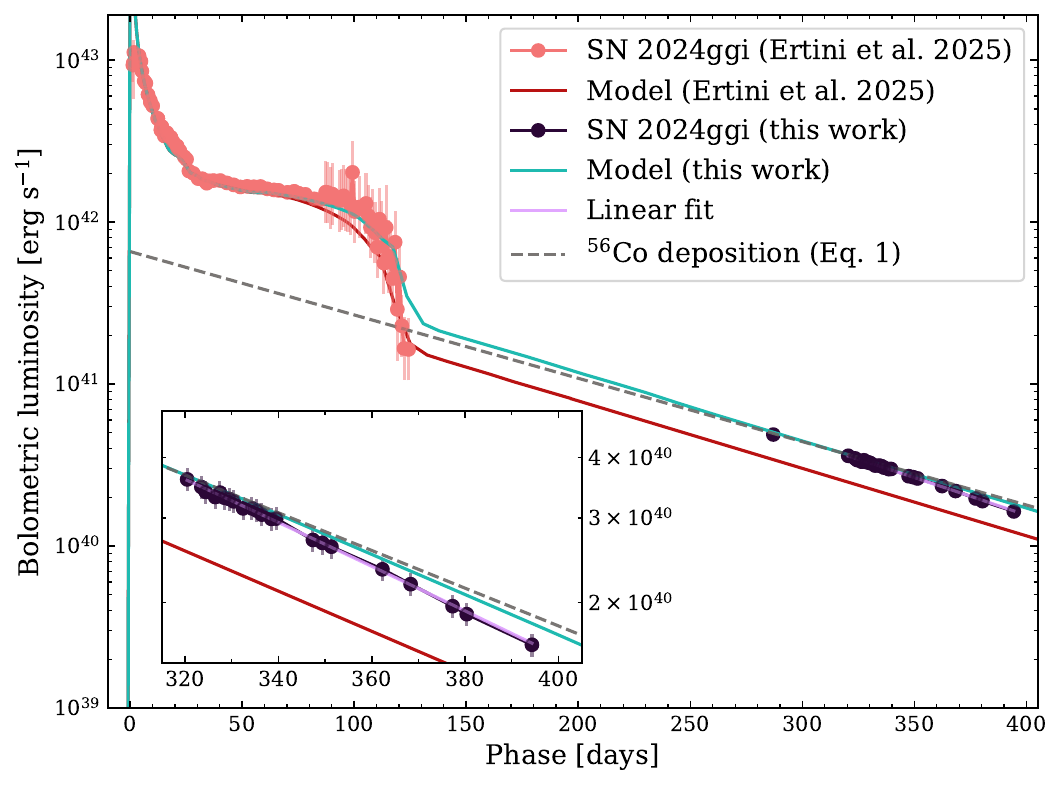}
    \caption{Bolometric light curve of SN 2024ggi. The bolometric light curve published in \citet{Ertini2025} and their preferred model are plotted in pink dots and line, respectively. Our fitting of Eq. \ref{eq:Ni_mass} is shown in purple. A new hydrodynamical model with the same parameters as those presented in \citet{Ertini2025}, but with a larger $^{56}$Ni mass of $0.052~M_\odot$ is shown in the cyan line. The dashed gray line indicates the decay with a $^{56}$Ni mass of $0.049~M_\odot$ and full-trapping of $\gamma$ rays.}
    \label{fig:bolometric}
\end{figure}

Finally, we computed the bolometric luminosity by integrating the SED over the observed Swope bands, adding the 60\%.
The resulting bolometric light curve is presented in Fig.~\ref{fig:bolometric}.
The decline rates were calculated with the same methodology used for the optical filters, and are also specified in Table \ref{tab:declines}. 
We do not note a significant break at 365 days, as noticed in the $ori$ bands.
Instead, the resulting slopes closely match the expected decline rate of the $^{56}$Co decay, and suggest a high efficiency in $\gamma$-ray trapping within the ejecta.

To estimate the synthesized $^{56}$Ni mass, we adopt the methodology of \cite{Hamuy2003}, which assumes complete trapping of $\gamma$-rays by the ejecta. This assumption may not hold if the ejecta becomes sufficiently diffuse at late times. To account for potential $\gamma$-ray escape, we incorporate the prescription proposed by \citet{Clocchiatti1997}, where the observed luminosity is decreased dependent on the $\gamma$-ray escape timescale $t_\gamma$.

Combining these approaches yields:

\begin{equation}\label{eq:Ni_mass}
    M_{Ni} = (7.866 \times 10^{-44})~\frac{L(t)}{1-\exp(-(t_{\gamma}/t)^2)}~\exp{\left[\frac{t - 6.1}{111.26}\right]} ~M_\odot ,
\end{equation}

\noindent where $L_t$ represents the bolometric luminosity (in erg s$^{-1}$) at the rest-frame epoch $t$.
We derived the $^{56}$Ni mass and gamma-ray escape timescale by fitting the bolometric light curve obtaining $\mathrm{M_{Ni}} = 0.049~M_\odot$ and $t_\gamma = 561$ days, respectively.
This high value is consistent with the decline rate of the bolometric light curve, confirming nearly complete $\gamma$-ray trapping. %in the ejecta.

A potential limitation of this method lies in the accuracy of the SED integration, particularly given the emission-line-dominated nature of the spectrum. To assess this, we compared the integration over the full spectral range with that of synthetic flux points from the observed spectra and found the discrepancy to be less than 5\%, confirming the robustness of our approach. 

As an additional validation, we used the pseudo-bolometric luminosity ($\mathrm{pL_{bol}}$) of SN 1987A as a reference, fitting the following relation:

\begin{equation}\label{eq:87A}
    \centering
    \frac{\mathrm{pL_{bol}}(t)}{\mathrm{pL_{bol87A}}(t)} = \left(\frac{M_{Ni}}{0.075}\right) \left( \frac{1-\exp(-(t_{\gamma}/t)^2)}{1-\exp(-(540/t)^2)}\right),
\end{equation}

\noindent where $t$ denotes the rest-frame time since explosion in days, and $t_\gamma$ represents the $\gamma$-ray escape timescale as in Eq.\ref{eq:Ni_mass} (see, e.g., \citealt{Spiro2014}). We adopted a redshift of z=0.001 for SN 1987A, while for SN 2024ggi we used the redshift value reported in Sect.~\ref{sec:observations}.
% \textbf{Cite source for 87A redshift}

We computed the pseudo-bolometric luminosity $\mathrm{pL_{bol}}(t)$ for SN 1987A by integrating its flux over the $BVRI$ bands \citep{Whitelock1988}.
To match the photometric systems of the observations of SN 1987A and SN 2024ggi, we performed synthetic photometry in the Johnson system using flux-scaled spectra of SN 2024ggi.
This allowed us to derive transformation factors between the Sloan and Johnson systems, which we subsequently applied to all photometric data. From this analysis, we obtained a $^{56}$Ni mass of $M_{\mathrm{Ni}} = 0.059~M_\odot$ and a $\gamma$-ray escape timescale of $t_\gamma = 553$ days for SN 2024ggi.

% Lastly, we compared the late-time bolometric light curve with a set of one-dimensional (1D) Lagrangian hydrodynamical models from \citet{Bersten2011}. Fig.~\ref{fig:bolometric} shows the model published by \citet{Ertini2025}, with an initial mass of 15 $M_\odot$ and a $^{56}$Ni mass of 0.035 $M_\odot$, and two models with higher $^{56}$Ni masses of 0.057 and 0.06 $M_\odot$. The lower-mass model is strongly incompatible with the bolometric light curve, while the other two show a better agreement.

% Finally, we compare the late bolometric light curve with the model presented in Ertini et al. (2025). This model was calculated using a one-dimensional (1D) Lagrangian hydrodynamic code (Bersten et al., 2011) that takes into account the deposition of gamma rays in the ejection assuming a grey approximation. It corresponds to a progenitor with an MZAMS of 15 M⊙, E = XXX, and a 56Ni mass of 0.035 M⊙ (pink line). However, this model clearly overestimates the 56Ni mass. We therefore calculate another model with the same parameters but with a different Ni mass and distribution in order to improve the comparison with the data. The new calculation, shown in green, corresponds to a higher Ni mass of 0.052 M_☉ and provides a much better match to the data. We note that previous estimates were based on photometry before the SN had actually reached the radioactive tail,  and this estimation was base on bolometric calculation provide by color calibration from Martinez et al. 2022 which are subject to large uncertainties.

Finally, we compared the late-time bolometric light curve with a suite of hydrodynamical models computed using the one-dimensional (1D) Lagrangian local thermodynamic equilibrium radiation hydrodynamics code from \citet{Bersten2011}. Fig.~\ref{fig:bolometric} displays the preferred model published by \citet{Ertini2025} (15 $M_\odot$ progenitor, $^{56}$Ni mass of 0.035 $M_\odot$) alongside a model with the same progenitor, explosion energy, and CSM parameters and a higher $^{56}$Ni mass ($0.052~M_\odot$). While the lower-mass model shows strong disagreement with the observed light curve, the second model provides a better match.

Our estimates, derived from three independent methods, are mutually consistent, resulting in a $^{56}$Ni mass of $0.05-0.06~M_\odot$, which is significantly higher than the 0.035 $M_\odot$, value reported by \citep[][the only published estimate to date]{Ertini2025}. This discrepancy likely stems from uncertainties in their bolometric light curve construction beyond 85 days, which relied solely on ($B-V$) color-based bolometric corrections from \citet{Martinez2022_CB}. Such an approach is subject to large uncertainties, while our approach is more robust. \citet{Dessart2025arXiv_ggi} propose a model with a $^{56}$Ni mass of $0.06~M_\odot$ that, despite not being fitted, reproduces the observed optical and NIR spectra at late times.

\section{Progenitor mass}
\begin{figure}
    \centering
    \includegraphics[width=.95\linewidth]{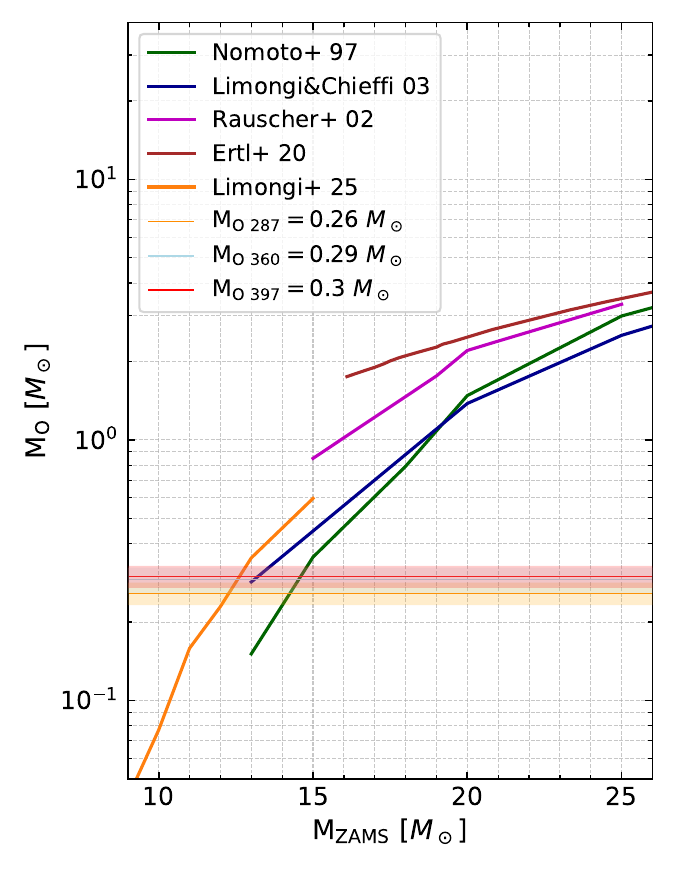}
    \caption{Oxygen core mass estimated from the measurement of the [\ion{O}{i}] doublet for phases 287, 360, and 396 days (horizontal lines). Oxygen production yields published by \citet{Nomoto1997}, \citet{Rauscher2002}, \citet{Limongi2003}, \citet{Ertl2020}, and \citet{Limongi2025arXiv} point to an initial mass of $12-15~M_\odot$ for the progenitor star.}
    \label{fig:O_mass}
\end{figure}

Nebular spectra are strongly dependent on the mass of the progenitor star. 
As a consequence, there are several methods to derive progenitor initial masses from nebular features.
Our first method resides in estimating an oxygen core mass from the [\ion{O}{i}]$~\lambda\lambda6300,6364$ luminosity, which is then linked to the progenitor mass through oxygen production yields \citep{Uomoto1986,Jerkstrand2014,Kuncarayakti2015}.
Nevertheless, a few caveats must be considered. This emission arises from the neutral oxygen, so the derived progenitor mass is a lower limit, and higher ionization levels must be assumed to be able to estimate a total oxygen mass. This procedure also depends strongly on the absolute flux calibration of the spectrum and on the estimate of the materials' temperature.
The [\ion{O}{i}]$~\lambda5577$ line is an indicator of this temperature (see e.g. \citealt{Jerkstrand2014}), but it is not recognizable in the spectra of SN 2024ggi presented here, as it appears blended with other emission lines.
Instead, we assume the material follows a temperature evolution similar to the model from \citet{Jerkstrand2014} for SN 2012aw.
In SN 2024ggi, the [\ion{O}{i}] doublet flux decreases slowly with time. The measured flux from the 287, 360, and 396 d spectra are $2.21\times10^{-13}$, $1.52\times10^{-13}$, and $1.15\times10^{-13}$ \fluxcgs, respectively. 
At the distance of the SN, this implies an oxygen mass of 0.3 $M_\odot$, which, when compared to theoretical oxygen production yields from \citet{Rauscher2002} and \citet{Ertl2020}, suggests a ZAMS mass well below the minimum mass included in their models of 15$M_\odot$. For those calculated by \citet{Nomoto1997} and \citet{Limongi2003}, the progenitor must have been a star between 12 and 15 $M_\odot$. Fig.~\ref{fig:O_mass} shows the cited production yields and our oxygen core mass estimations for the three epochs.

The [\ion{O}{i}]/[\ion{Ca}{ii}] ratio has been suggested as a more reliable indicator of the progenitor mass \citep{Fransson1989,fang2022}, as it does not depend on the absolute flux calibration of the spectrum.
Moreover, this value does not change significantly during the evolution, making this method less sensitive to the observation epoch. 
In addition, the ratio is apparently unaffected by possible interaction between the SN ejecta and surrounding material \citep{Folatelli2025}.
The progenitor mass can be constrained by comparing the flux ratio with those from synthetic spectra.
In SN 2024ggi, the measured ratios were $0.48\pm0.07$, $0.49\pm0.09$, and $0.48\pm0.07$ for each epoch, respectively.
Compared with models from \citet{Jerkstrand2014} and \citet{Dessart2021}, these values correspond to a somewhat low mass progenitor, around $10 - 12~M_\odot$.

\begin{figure}
    \centering
    \includegraphics[width=0.9\linewidth]{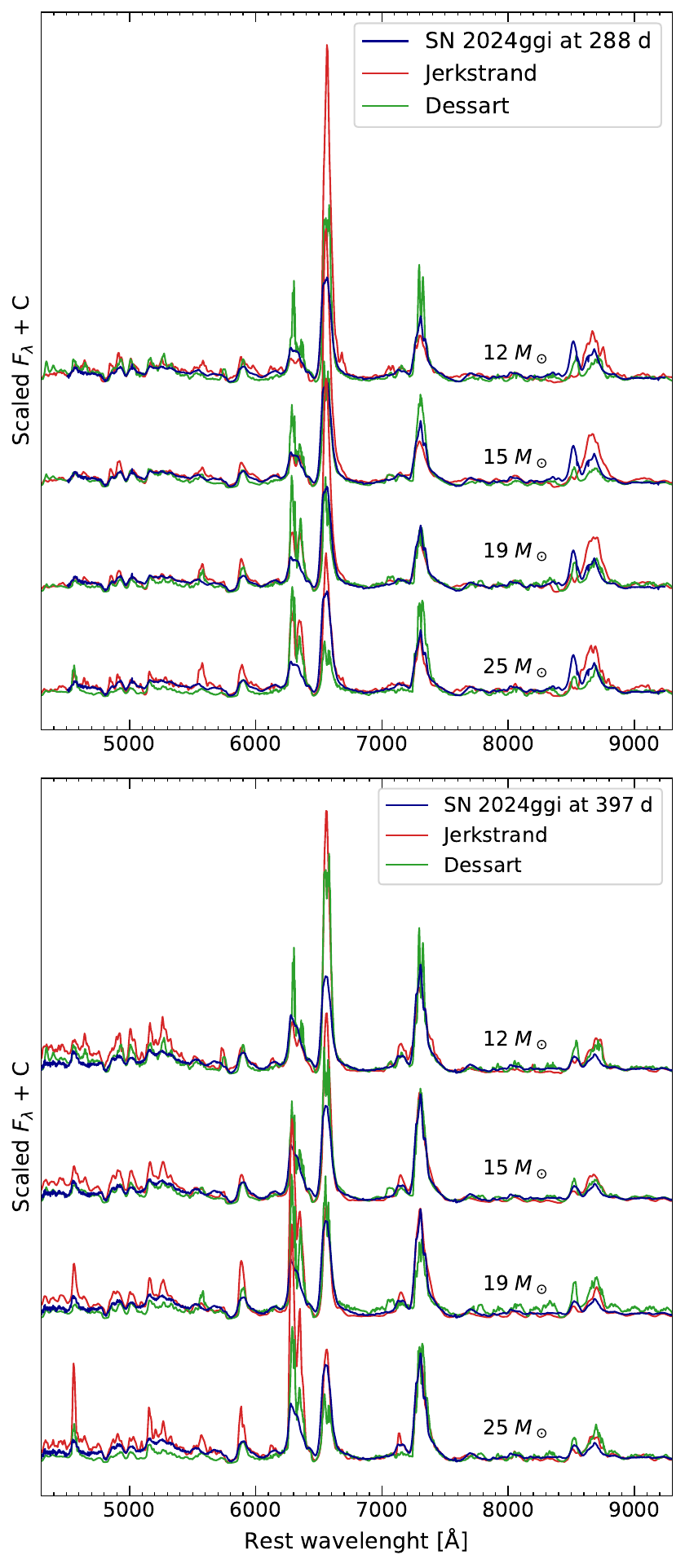}
    \caption{SN 2024ggi in comparison with synthetic spectra from \citealt{Jerkstrand2014} and \citealt{Dessart2021}. All spectra are scaled by the mean value between $6900-8200$ \AA\ as an indication of the continuum. The upper plot shows the spectra at 288 days, while the synthetic spectra phases are 360 and 250 days for \citealt{Dessart2021} and \citealt{Jerkstrand2014} models, respectively. In the bottom plot, we show our last spectrum at 396 days, this time changing spectra from \citealt{Jerkstrand2014} with those at the phase of 400 days.}
    \label{fig:models}
\end{figure}

% wgEqJNhJm5jGtEp
Another way of arriving at a mass estimation is by directly comparing the observed spectra with synthetic spectra from the literature, such as those published by \cite{Dessart2021} and \cite{Jerkstrand2012}. Fig.~\ref{fig:models} shows the comparison of our observed spectra with the series of model spectra computed for varying initial progenitor masses at similar ages.
The models that better match the SN emission lines correspond to relatively low masses, of 12 and 15 $M_\odot$.
Higher-mass models significantly over-produce the oxygen and sodium features.
 % \textbf{Add 2-4 more sentences detailing more exactly what you mean by “better match” and “differ significantly}
%We notice that the H$\alpha$ feature is over-estimated by the selected models. ... \textbf{continue} 

\label{sec:prog_mass}
% \begin{figure}
%     \centering
%     \includegraphics[width=0.85\linewidth]{figures/24ggi_Jerkstrand_norm_Ca.png}
%     \caption{Caption}
%     \label{fig:enter-label}
% \end{figure}

\section{Conclusions}\label{sec:conclusion}

We present optical photometric and spectroscopic observations of the nearby Type II SN 2024ggi during its nebular phase. The dataset includes four spectra obtained at 287, 288, 360, and 397 days after explosion, along with $uBVgri$ photometry spanning days 320 to 400. The observations were conducted at the Gemini Observatory and Las Campanas Observatory. This comprehensive dataset enabled us to investigate both the explosion properties and the nature of the progenitor star, yielding a progenitor mass estimate that is independent from those previously published.

The nebular spectra display the characteristic emission features of Type II SNe, with prominent forbidden lines of oxygen, calcium, and (semi-forbidden) magnesium, as well as strong H$\alpha$ and the \ion{Ca}{ii} infrared triplet. While most emission features exhibit single-peaked, symmetric profiles, the [\ion{Fe}{ii}] $\lambda$7155 line shows a distinct double-peaked structure. The line profiles overall suggest a roughly spherical ejecta geometry, with minimal Doppler shifts and little temporal evolution. No signatures of late-time interaction with circumstellar material, such as variability in H$\alpha$ or excess emission on the blue part of the spectrum, are detected up to 400 days post-explosion.

The late-time light curves decline at a rate of approximately 1 mag per 100 days across all filters, consistent with the expected behavior of Type II SNe. A notable break appears in the redder bands around day 365, potentially indicating a decline in the strength of H$\alpha$ and calcium features relative to the continuum. Using near-infrared data (Mera et al., in prep), we constructed the bolometric light curve and inferred a $^{56}$Ni mass of $\approx 0.05-0.06~M_\odot$, significantly higher than the 0.035 $M_\odot$ reported by \citet{Ertini2025}. We attribute this discrepancy to the large uncertainties in the methodology for computing the second half of the bolometric light curve in that work.

To constrain the progenitor mass, we compared our spectra with nebular models from \citet{Jerkstrand2014} and \citet{Dessart2021}. Both direct spectral comparisons and the [\ion{O}{i}]/[\ion{Ca}{ii}] flux ratio suggest a progenitor initial mass of $10-15~M_\odot$.
From the luminosity of the [\ion{O}{i}] line alone, we estimate an oxygen core mass of $0.26-0.30~M_\odot$, which corresponds to an initial mass below 15 $M_\odot$ based on oxygen yields from \citet{Rauscher2002} and \citet{Ertl2020}. Using nucleosynthesis models from \citet{Nomoto1997}, \citet{Limongi2003}, and the more recent yields from \citet{Limongi2025arXiv}, we refine this to a progenitor with $M_{\mathrm{ZAMS}}$ between 12 and 15 $M_\odot$. 
%All three methods yield consistent results, indicating that SN 2024ggi originated from a star in this mass range.

% We make use of nebular synthetic spectra from \citet{Jerkstrand2014} and \citet{Dessart2021} to arrive at an initial mass of the SN progenitor. Performing a direct comparison of the spectra with the models and the [\ion{O}{i}] to [\ion{Ca}{ii}] ratio yields a progenitor mass between 10 and 15 $M_\odot$. On the other hand, analyzing the [\ion{O}{i}] emission alone, we derive an oxygen core mass of $M_O = 0.26-0.3 ~ M_\odot$. Using the oxygen production yields from \citet{Rauscher2002} and \citet{Ertl2020}, this translates to a star with an initial mass well below $15~M_\odot$. Following the nucleosynthesis from \citet{Nomoto1997} and \citet{Limongi2003}, the progenitor mass must have been a star of $13 - 15~M_\odot$. More recent work by \citet{Limongi2025arXiv} calculates the ion production in the lower end of the mass distribution. With their results, we find that the progenitor had $12-13~M_\odot$. Our three methods yield consistent results, concluding that the progenitor star of SN 2024ggi had a $M_{ZAMS} = 12-15~M_\odot$.

Our progenitor mass estimate is consistent with previous studies. \citet{Xiang2024} report a mass of $13 \pm 1~M_\odot$, and \citet{Ertini2025} suggest 15 $M_\odot$, both consistent with our results. While \citet{Hong2024} finds a slightly lower mass of 10.2 $M_\odot$, the value derived by \citet{Chen2024} is slightly higher, in the range of $14-17~M_\odot$.
Further optical and infrared observations will be essential to study potential dust formation and the eventual onset of interaction between the ejecta and the surrounding CSM.

% This result is in agreement with the previous estimates for SN 2024ggi published by \citet{Xiang2024}, of $13 \pm 1~M_\odot$, and by \citet{Ertini2025}, of $15~M_\odot$.
% The initial mass derived by \citet{Hong2024} of $10.2~M_\odot$ falls a bit below our estimate, while \citet{Chen2024} findings are slightly higher, of $14-17~M_\odot$. 

% The archival image analyses showed a progenitor candidate consistent with an RSG star.
% \citealp{Xiang2024} combined Hubble Space Telescope (HST) and 
% Spitzer Space Telescope pre-explosion images to characterize the progenitor as a bright, red star with an estimated initial mass of $13 \pm 1~M_\odot$. 
% \citealp{Chen2024} independently proposed a slightly more massive progenitor ($14-17$ $M_\odot$) using Dark Energy Camera Legacy Survey data.
% An alternative approach by \citealt{Hong2024} leveraged HST imaging to analyze the local stellar environment. By dating the youngest stellar population near the supernova site, they inferred a progenitor mass of 10.2 $M_\odot$, providing a complementary estimate to direct detection methods.
% Lastly, \citealt{Ertini2025} presented the first hydrodynamical model of the bolometric light curve during the full plateau phase and derived a progenitor mass of $15~M_\odot$.

\begin{acknowledgements}
L.G. acknowledges financial support from AGAUR, CSIC, MCIN and AEI 10.13039/501100011033 under projects PID2023-151307NB-I00, PIE 20215AT016, CEX2020-001058-M, ILINK23001, COOPB2304, and 2021-SGR-01270.

K.M. acknowledge support from NASA grants JWST-GO-02114,
JWST-GO-02122, JWST-GO-04522, JWST-GO-04217, JWST-GO-04436,
JWST-GO-03726, JWST-GO-05057, JWST-GO-05290, JWST-GO-06023,
JWST-GO-06677, JWST-GO-06213, JWST-GO-06583. Support for
programs \#2114, \#2122, \#3726, \#4217, \#4436, \#4522,  \#5057,
\#6023, \#6213, \#6583, and \#6677
were provided by NASA through a grant from the Space Telescope Science
Institute, which is operated by the Association of Universities for Research in
Astronomy, Inc., under NASA contract NAS 5-03127.

% W.B.H. NSF
Some of this material is based upon work supported by the National Science Foundation Graduate Research Fellowship Program under Grant Nos. 1842402 and 2236415. Any opinions, findings, conclusions, or recommendations expressed in this material are those of the author(s) and do not necessarily reflect the views of the National Science Foundation. 

M.D. Stritzinger is funded by the Independent Research Fund Denmark (IRFD, grant number  10.46540/2032-00022B) and by an Aarhus University Research Foundation Nova project (AUFF-E-2023-9-28).
\end{acknowledgements}

\bibliographystyle{aa_url} % style aa.bst
\bibliography{biblio} % your references Yourfile.bib

%\begin{thebibliography}{}
%
%  \bibitem[Baker(1966)]{baker} Baker, N. 1966,
%      in Stellar Evolution,
%      ed.\ R. F. Stein,\& A. G. W. Cameron
%      (Plenum, New York) 333

%\end{thebibliography}

\end{document}